\documentclass[a4paper]{jpconf}
\usepackage{graphicx}
\usepackage{xcolor}
\usepackage{cite}
\begin{document}
\title{1D valence bond solids in a magnetic field}

\author{Adam Iaizzi and Anders W. Sandvik}

\address{Department of Physics, Boston University, 590 Commonwealth Avenue, Boston, Massachusetts 02215, USA}

\begin{abstract}
A Valence bond solid (VBS) is a nonmagnetic, long-range ordered state of a quantum spin system where local spin singlets are formed 
in some regular pattern. We here study the competition between VBS order and a fully polarized ferromagnetic state as function of an external 
magnetic field in a one-dimensional extended Heisenberg model---the J-Q$_2$ model---using stochastic series expansion (SSE) quantum Monte 
Carlo simulations with directed loop updates. We discuss the ground state phase diagram.
\end{abstract}
\section{Introduction}
A valence bond solid (VBS) is a long-range nonmagnetic ordered state that can appear in certain quantum spin systems. 
In a VBS (also called the spin-Peierls state) spins spontaneously pair up to form singlets in some regular pattern, breaking lattice translational symmetry but retaining spin rotation symmetry. 
Recent innovations in models and simulation techniques have enabled large-scale numerical studies of this type of ground state and associated quantum phase transitions. 
In one dimension, these studies have found evidence for the fractionalization of triplons into deconfined spinons carrying spin $S=1/2$ in the VBS phase~\cite{tang2011a}. 
Since lattice translational symmetry is discrete, the VBS transition is allowed in one-dimensional systems at zero temperature. 
We here add an external magnetic field and study its effect on the VBS. 

The J-Q model is at its core a sign-problem-free system for studying valence bond solids, associated quantum phase transitions, and other related phenomena using quantum Monte Carlo (QMC) methods~\cite{Sandvik2007}. 
It supplements a standard Heisenberg $J$ exchange term with a multispin interaction composed of a product of two (or more) singlet projection operators, 
\begin{equation}
P_{ij} = \frac{1}{4} - \vec S_i \cdot \vec S_j. 
\end{equation}
Several variants of the J-Q model have been studied extensively in two dimensions~\cite{Sandvik2007, sandvik2011, lou2009, jin2013, kaul2013, kaul2014}. 
The J-Q$_2$ model has a Q term with two singlet projection operators; its one-dimensional realization can be written as
\begin{equation}
H \equiv -J \sum \limits_{i} P_{i,i+1} - Q \sum \limits_{i} P_{i,i+1} P_{i+2,i+3} + h \sum \limits_i S^z_i
\end{equation}
where $J,Q > 0$ indicate antiferromagnetic interactions. 
In this paper we will fix $J=1$ and use the dimensionless parameters $q \equiv Q/J$ and $h \equiv h_{bare} /J$. 
At zero temperature and zero field the 1D J-Q$_2$ model produces a VBS when $q$ exceeds  $q_c=(Q/J)_c=0.84831$~\cite{tang2011a}.


\section{Methods}

The primary computational tools we are using in this work are exact diagonalization of the Hamiltonian for small systems and a stochastic series expansion (SSE) QMC method for larger systems. 
Our QMC simulations are based on the directed loop algorithm for the anisotropic Heisenberg model in an external field~\cite{sandvik_dl}, with added procedures to account for the Q term. 
Strictly speaking, this is a finite temperature method, but since any finite size system will have a finite excitation gap, we can reach the zero temperature regime by using a 
temperature much smaller than the gap. 

\section{Results}

\begin{figure}[h]
\includegraphics[width=20pc]{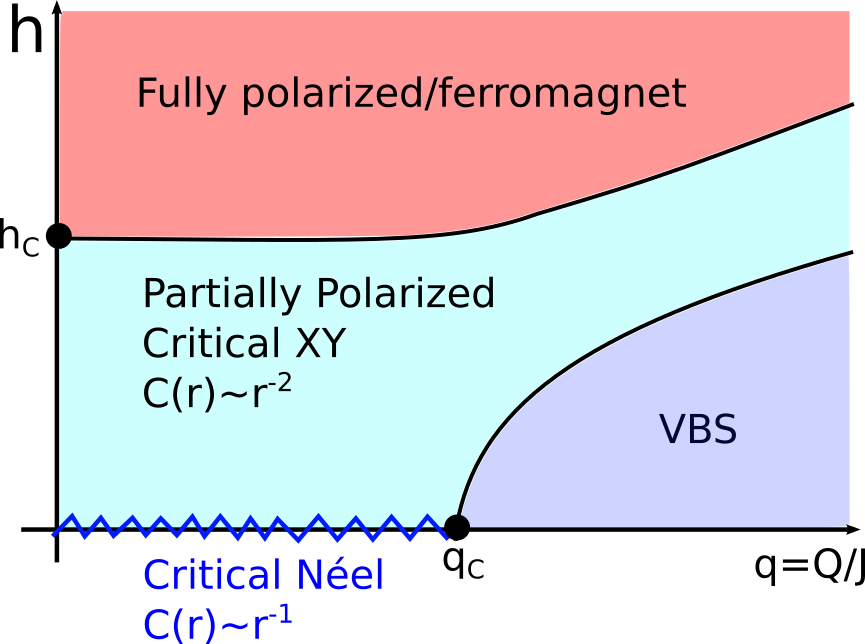}\hspace{2pc}%
\begin{minipage}[b]{14pc}\caption{\label{pd}Schematic preliminary phase diagram of the 1D JQh model in the plane of the coupling ratio $q$ and magnetic field $h$. }
\vspace{5pc}
\end{minipage}
\end{figure}

We begin by showing in Fig.~\ref{pd} a schematic phase diagram based on previous work and our own calculations (discussed in more detail below).
Along the $h$-axis, with $q=0$, we have the standard Heisenberg chain in an external field, which undergoes a transition to a fully polarized ferromagnet at $h=h_c$. 
Along the $q$-axis for $h=0$ we have the previously studied zero-field J-Q$_2$ chain, which for $q<q_c$ has a quasi-N\'{e}el ordered state with spin correlations decaying with distance $r$ as $1/r$ (up to logarithmic corrections)~\cite{singh1989}, and  at $q=q_c$ undergoes a phase transition to a VBS~\cite{tang2011a,sanyal2011}. 
This transition is of the same type (similar to a Kosterlitz-Thouless transition) as in the the well-studied frustrated J$_1$-J$_2$ chain. 

\begin{figure}[h]
\begin{minipage}{18pc}
\includegraphics[width=18pc]{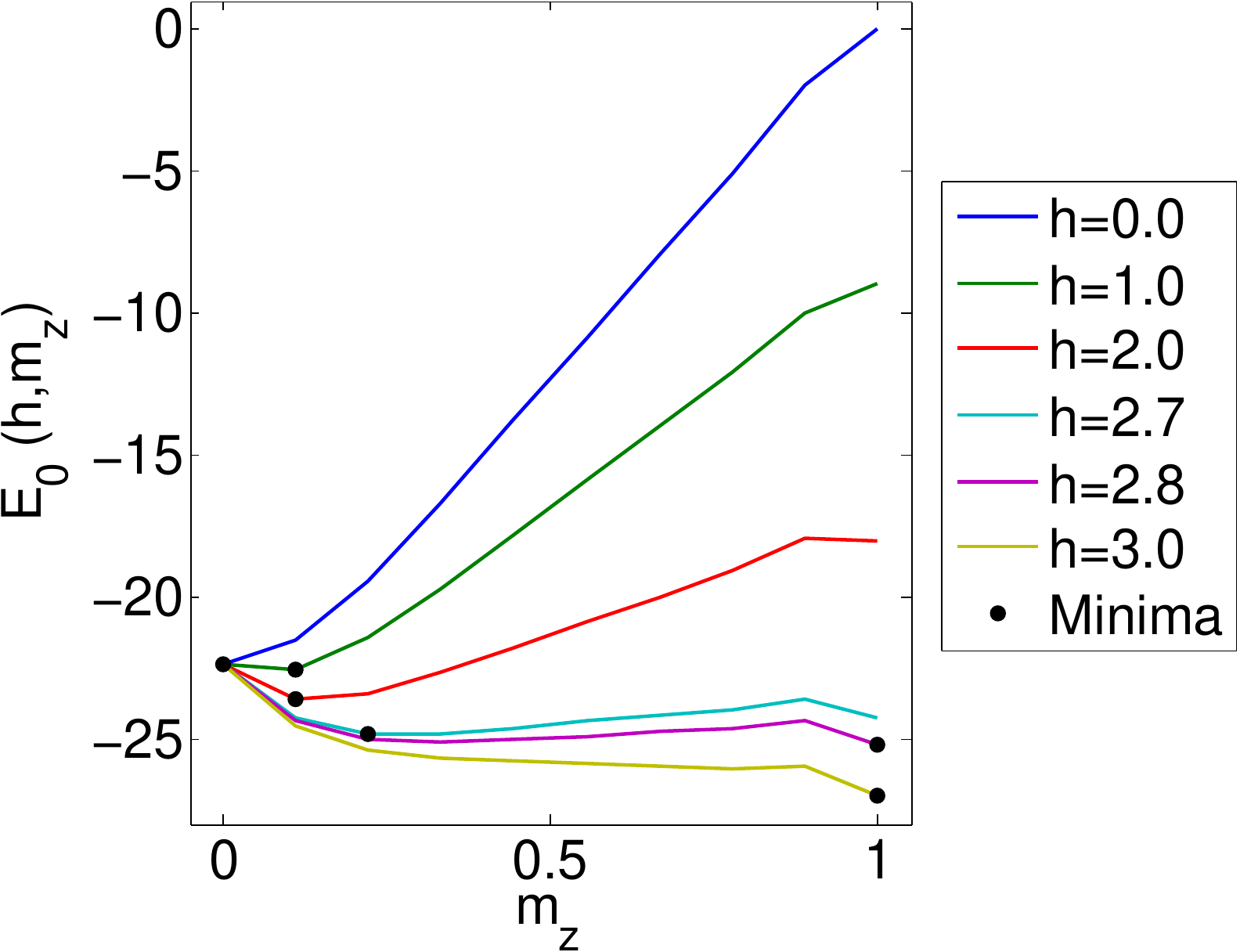}
\caption{\label{en-v-mz} Ground state energy as a function of magnetization for different values of $h$ showing either side of the magnetization jump. Computed using exact diagonalization for $L=18$ and $q=1.2$ with open boundary conditions. }
\end{minipage}
\hspace{2pc}
\begin{minipage}{18pc}
\includegraphics[width=18pc]{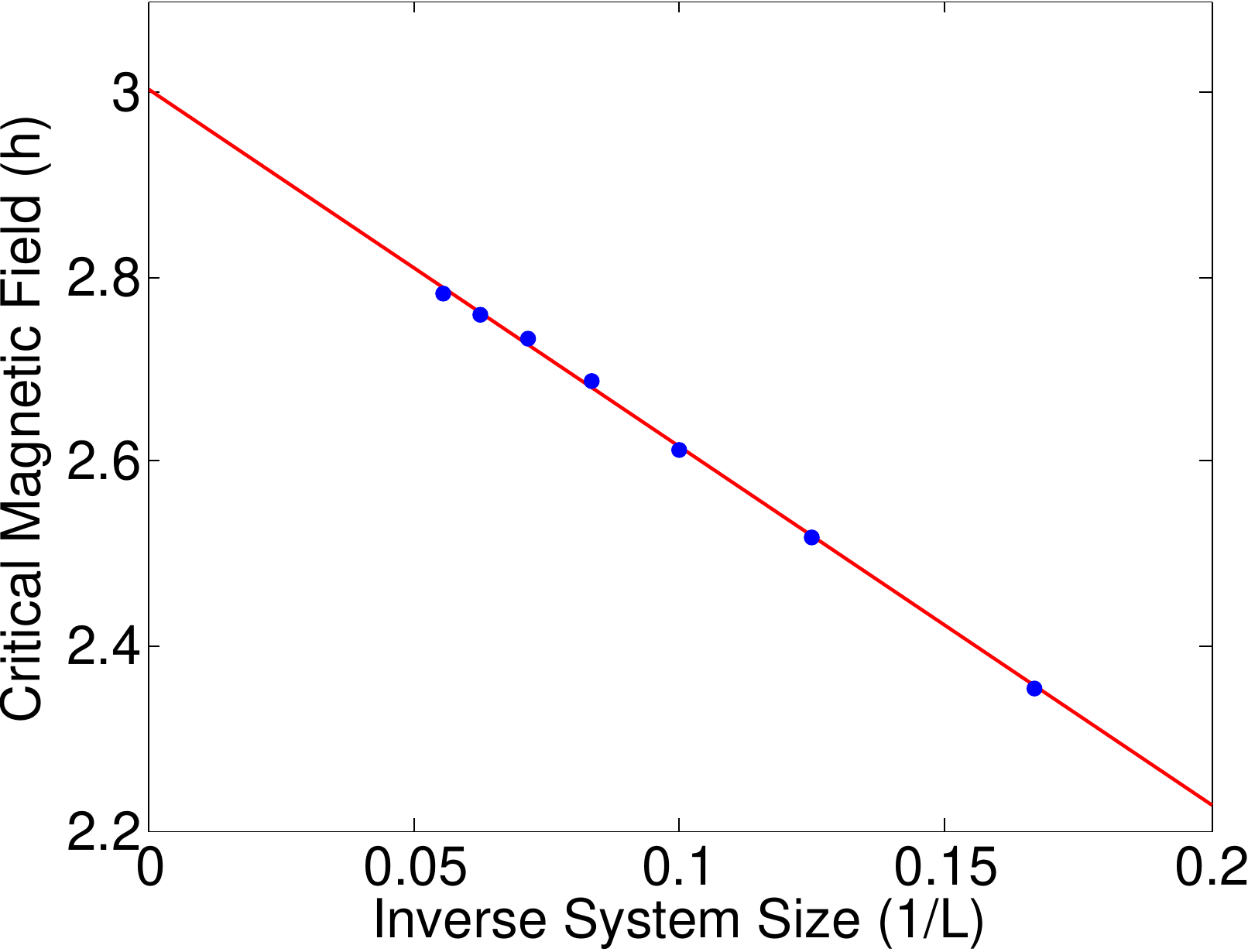}
\caption{\label{hcrit} Finite size scaling: critical magnetic field for $q=1.2$ plotted against $1/L$. Plotting against inverse system size we can extract $h_c$ in the thermodynamic limit by taking the $y$-intercept of a linear fit. We find $h_c (L\rightarrow \infty) \approx 3$ with $R^2>0.999$. Computed using exact diagonalization with open boundary conditions. }
\end{minipage}
\end{figure}

The area off of the axes of Fig.~\ref{pd} has not been studied previously. 
We believe there are three different phases: a fully polarized state, a partially polarized state with critical XY spin correlations (like the Heisenberg chain in a magnetic field) and a VBS state. 
The critical magnetic field, $h_c (q)$, can be quantitatively extracted from data such as those shown in Fig.~\ref{hcrit}. 
In some cases we see clear magnetization jumps, i.e. a first-order transition, but it remains an open question whether the transition is first-order along the whole line, or only above some minimum value of $q$. 

The $q_c$ line extends out from the known zero-field transition point; the rest of the line depends on how the VBS is destroyed by the magnetic field. 
The elementary excitations of the $h=0$ VBS in one dimension are pairs of deconfined spinons.
Each pair of spinons arises from breaking one of the singlet bonds, creating a triplet which fractionalizes into two independently propagating domain walls between the two possible VBS ordering patterns. 
We believe that an arbitrarily low density of such deconfined spinons will destroy the VBS state (using arguments analogous to the standard ones for breaking of a discrete symmetry by temperature in one dimension, e.g., in the Ising chain).
Thus, the magnetic field required to destroy the VBS should be simply related to the spin gap.

\begin{figure}
\begin{minipage}{18pc}
\includegraphics[width=18pc]{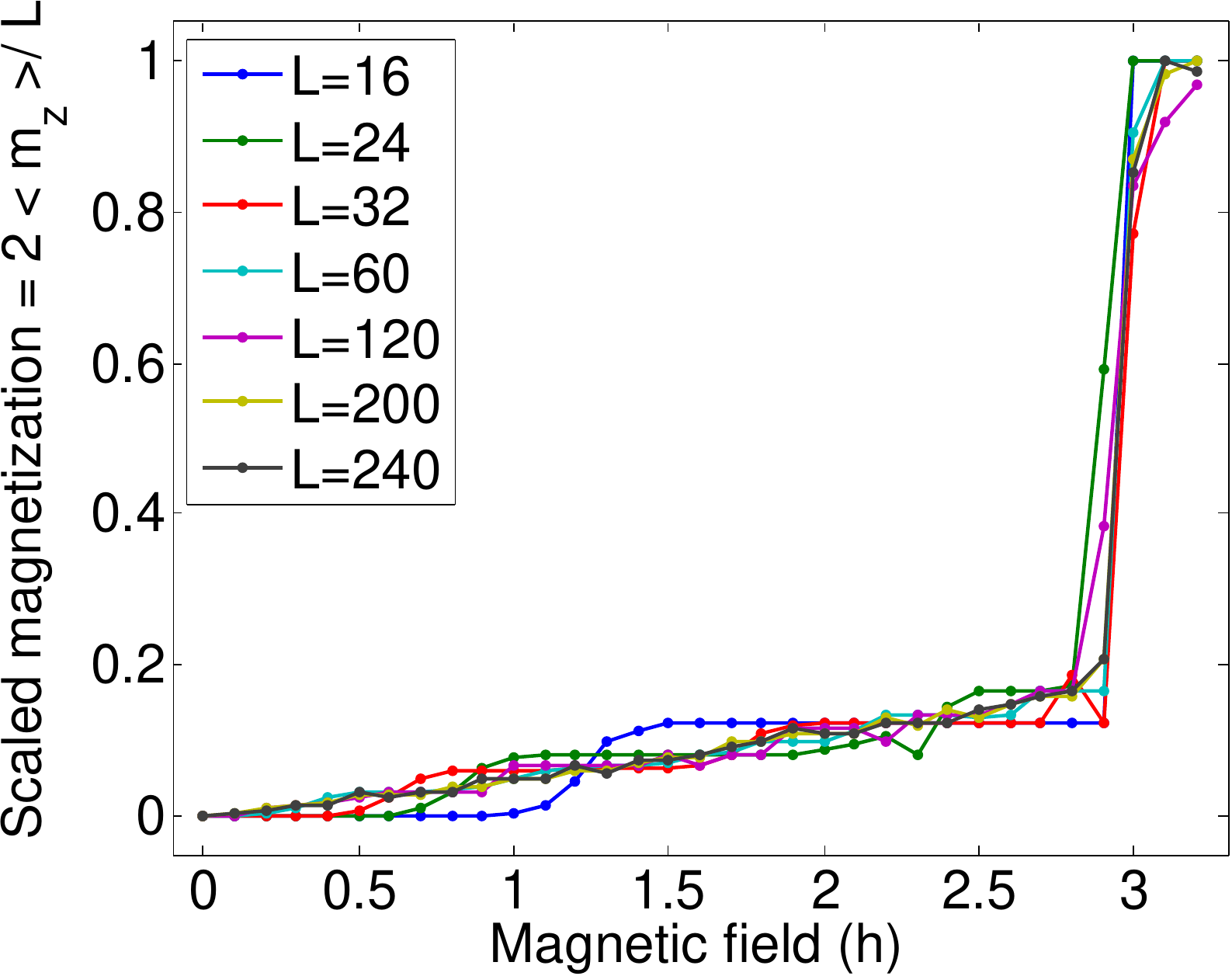}
\caption{\label{magvh}Scaled magnetization plotted as a function of applied magnetic field with $q=1.2$ for a wide range of sizes. Computed using QMC with periodic boundary conditions.}
\end{minipage}
\hspace{2pc}
\begin{minipage}{18pc}
\includegraphics[width=18pc]{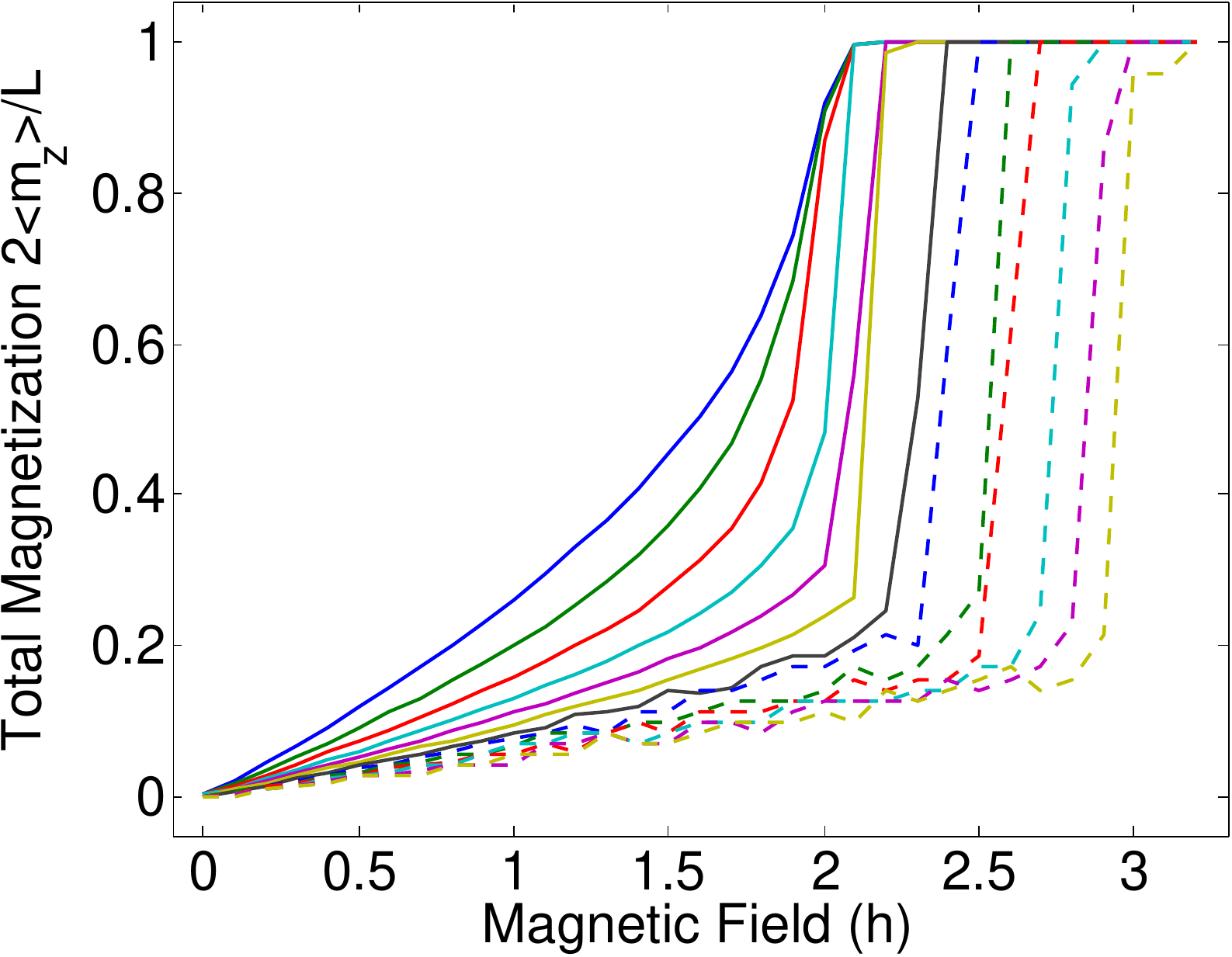}
\caption{\label{varQ} Scaled magnetization plotted as a function of applied magnetic field for a range of different values of $q=Q/J$. From the left (solid blue), $q=0.0, \, 0.1, \, 0.2... \, 1.2$. Computed using QMC with $L=140$ and open boundary conditions. }
\end{minipage}
\end{figure}

Starting in the VBS phase ($q=1.2$) and increasing the magnetic field we observe a jump in the magnetization. 
In Fig.~\ref{magvh} we plot the scaled magnetization, 
\begin{equation}
\left< m \right>= \frac{2}{L} \left| \sum S^z_i \right|
\end{equation}
for $0 \leq h \leq 3.2$ for system sizes from $L=16$ to $L=240$.
At every size we find the same behavior: roughly linear response until $m\approx 1/6$ followed by a sharp jump to the fully polarized state. 
This resembles the so-called metamagnetic transition demonstrated in the anisotropic J$_1$-J$_2$ chain~\cite{gerhardt1998,hirata1999,aligia2000,dmitriev2006}. 

In Fig.~\ref{varQ} we observe the magnetization jump developing from $q=0$, the Heisenberg limit. 
For $q=0$, the ground state becomes fully polarized at $h \approx 2$. 
There is no jump and $m(q,h)$ should be continuous in the thermodynamic limit. 
For small values of $q > 0$, $m(h)$ becomes steeper, but based on this data we cannot determine if there is a jump in the magnetization until larger values of $q$. 
As $q$ is increased the jump becomes more distinct and the maximum magnetization before the jump asymptotically approaches $m_c(q) \rightarrow 1/6$. 
Fig.~\ref{varQ} was generated using open boundary conditions in an attempt to `pin' the structure of the ground state right before the magnetization jump. 
No significant differences were observed in large systems between results generated with open and periodic boundary conditions. 


To get a better idea of the behavior around this phase transition, we have solved small chains using exact diagonalization. 
In Fig.~\ref{en-v-mz} we plot ground state energy of an 18 site chain for each magnetization sector, $E_0 (h,m) = E_{min}(q,m) - h m$, for a selection of magnetic fields. 
For each magnetic field, the black dot shows which value of $m$ is the ground state. 
For $h=0$, we can see that there is a slight change in curvature at the high magnetization states; this is key to the transition because it allows the fully polarized state to `outrun' intermediate polarizations when $h$ is increased. 
A similar change in curvature was observed in the metamagnetic transition in the J$_1$-J$_2$ model~\cite{hirata1999}.
This change in curvature appears only for $q>0$; it is not a feature of the Heisenberg model.
For $h=2.7$,  the ground state has a magnetization of $m=0.22$; at just a slightly higher field, $h=2.8$, the ground state is fully polarized.
The ground state jumps from $m=0.22$ to $m=1$ without passing though any of the intermediate states.

We can also use exact diagonalization to examine the critical magnetic field; in Fig.~\ref{hcrit} we plot $h_c (q=1.2,1/L)$ for $L=6$ to $L=18$. From this we obtain a good linear fit which predicts that $h_c (L\rightarrow \infty)=3$ with $R^2 > 0.999$, which is consistent with our QMC data. By repeating this process over a mesh of $q$ values we find a line consistent with the qualitative picture presented in Fig.~\ref{pd}. 

\begin{figure}
\includegraphics[width=18pc]{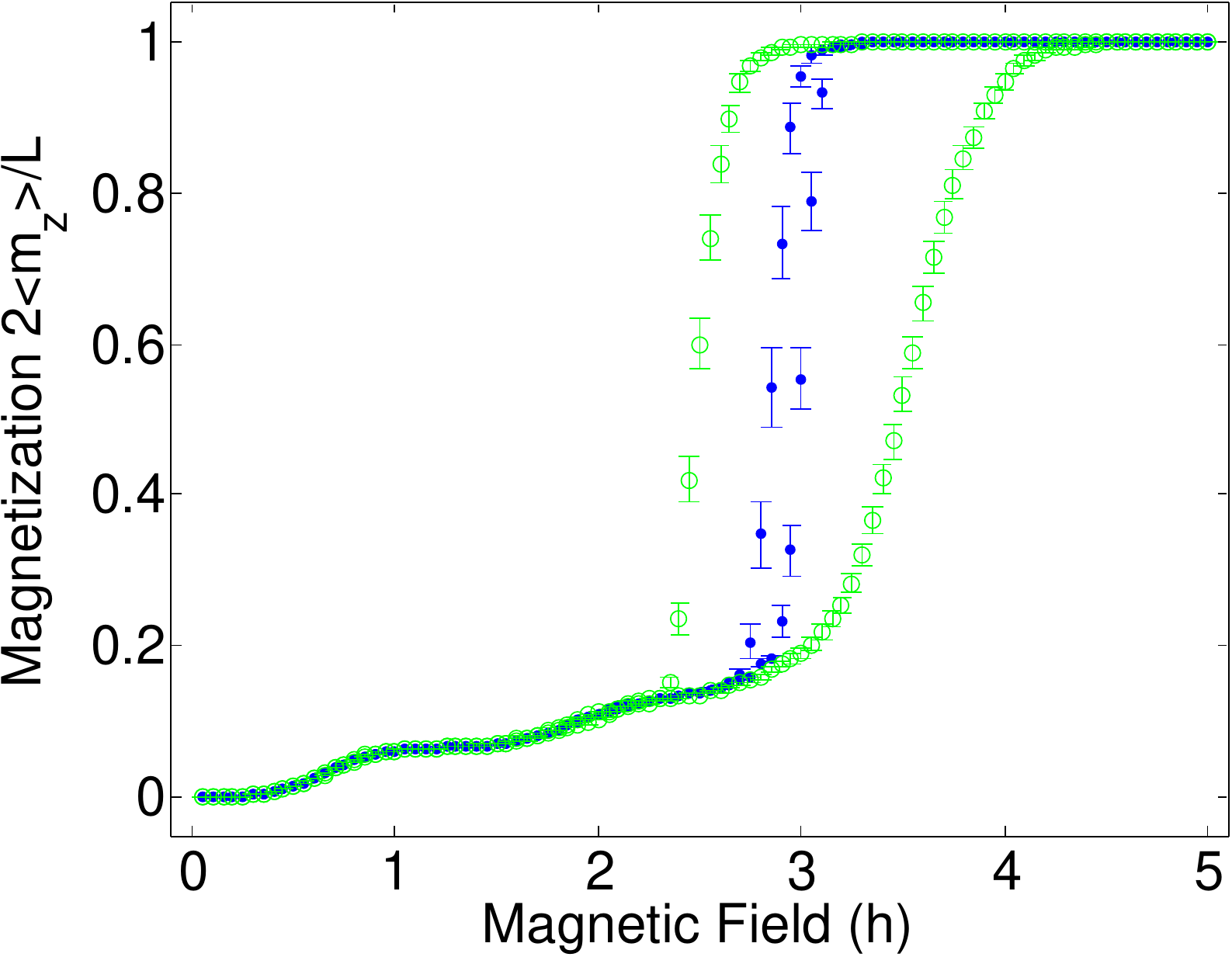}\hspace{2pc}%
\begin{minipage}[b]{18pc}\caption{\label{hloop} Simulation time hysteresis for a $L=30$ chain with $q=1.2$ and $\beta = 8$. $\textcolor{blue}{\fullcircle}$ corresponds to a field velocity of $v_1 =  2.5 \times 10^{-7}$ in units of magnetic field per Monte Carlo sweep, $\textcolor{green}{\opencircle}$ has a faster velocity of $v_2 = 10^{-5}$. Computed using QMC data with periodic boundary conditions. }
\vspace{4pc}
\end{minipage}
\end{figure}

In Fig.~\ref{hloop}, we present QMC data for $q=1.2$ and $L=30$. The system exhibits hysteresis when the magnetic field is ramped up and back down, a hallmark behavior in QMC simulations of a first-order phase transition. 
For clarity, only two different sweep velocities are shown here; a finer mesh of sweep velocities revealed a consistent positive relationship between velocity and the size of the hysteresis effect. 

\section{Discussion}

The magnetization jumps and hysteresis lead us to believe that the J-Q$_2$ model produces a first-order transition from a partially magnetized to a fully polarized state. 
The mechanism for this transition is currently unknown and is the subject of an ongoing investigation. We suspect it may bear resemblance to the metamagnetic transition. 

We draw these preliminary conclusions from limited data. 
At low temperature and high values of $q$ and $h$ our QMC method encounters difficulty reaching equilibrium, especially around the magnetization jump. 
This often causes simulations to become stuck in metastable magnetization states and produce poor estimates of statistical error. 
As a result, the magnetization curves in Figs. \ref{magvh} and \ref{varQ} are not very smooth. 

We are currently developing a replica exchange method (also known as ``quantum parallel tempering") we hope will ameliorate this problem. 
Replica exchange was developed for simulating classical systems with slow relaxation times~\cite{hukushima1996} and has since been adapted for use with quantum systems~\cite{sengupta2002}. 
Despite equilibration issues, we are reasonably confident in our preliminary results and the conclusions we have drawn from them, since repeated independent simulations produce data that match to within small statistical noise. 

This project is part of a larger plan to study the J-Q-h model in two dimensions. 
The simulation methods developed for one-dimensional systems should work with few modifications in two dimensions. 

\ack
We would like to thank Kedar Damle for valuable discussions facilitated by the APS-IUSSTF Physics PhD Student Visitation Program. 
This work was supported by the NSF under grants No.~DMR-1104708 and DMR-1410126.

\section*{References}

\bibliography{bibstuff.bib}

\end{document}